\documentclass[
aps,
prb,
reprint,
showkeys,
twocolumn,
nofootinbib,
superscriptaddress
]{revtex4-2}

\usepackage{natbib}
\usepackage{amssymb}
\usepackage{color}
\usepackage{amsfonts}
\usepackage{textcomp}
\usepackage{amsmath}
\usepackage[charter]{mathdesign}
\usepackage{easyReview}
\usepackage{microtype} 
\usepackage{upgreek}
\usepackage{booktabs}

\definecolor{link}{RGB}{57,106,177}
\definecolor{blue}{RGB}{0,0,0}
\definecolor{darkgreen}{RGB}{0,128,0}
\usepackage{float}
\usepackage{graphicx}
\usepackage{hyperref}
\hypersetup{
	colorlinks=true,
	linkcolor=link,
	citecolor=link,
	urlcolor=link
}

\begin{document}

\title[Andreev-reflection spectroscopy of SrPt$_3$P single crystals]{Andreev-reflection spectroscopy of SrPt$_3$P single crystals}
\author{Valeriy A. Stepanov}
\affiliation{P. N. Lebedev Physical Institute, Russian Academy of Sciences, 119991 Moscow, Russia}
\author{Nikolai D. Zhigadlo}
\affiliation{CrystMat Company, CH-8037 Zurich, Switzerland}
\author{Erik Piatti}\email{erik.piatti@polito.it}
\author{Dario Daghero}\email{dario.daghero@polito.it}
\affiliation{Department of Applied Science and Technology, Politecnico di Torino, 10129 Torino, Italy}%


\begin{abstract}
The interplay between inversion symmetry breaking and superconducting pairing mechanisms is a long-standing puzzle of solid-state physics.
Ternary platinum phosphide SrPt$_3$P is an excellent case study in this regard, since it shares the distorted anti-perovskite crystal structure of non-centrosymmetric superconductor CePt$_3$Si but preserves the inversion symmetry center, enabling a direct comparison between the two compounds.
Despite intense scrutiny, the superconducting pairing strength, gap symmetry, and magnitude of SrPt$_3$P remain an open question.
Here, we study the gap structure of high-quality SrPt$_3$P single crystals with $T_c = 8.6$\,K by using point-contact Andreev-reflection spectroscopy.
The spectra systematically showed symmetric conductance maxima and a zero-bias minimum, as is typical of fully gapped superconductors, and dips at finite voltage $V_\mathrm{dip}$ due to the local transition of the contact region to the resistive state.
In contacts with high resistance, where the dips fall sufficiently far from the gap features, the conductance curves can be well fitted with a single isotropic $s$-wave gap.
The temperature dependence $\Delta(T)$ of the gap amplitude extracted from the fit is in good agreement with the one predicted by the BCS theory, with a zero-temperature mean value $\Delta(0)=1.88\pm0.08$\,meV, and a mean gap ratio $2\Delta/k_\mathrm{B}T_c = 5.2\pm0.2$.\\\\
Cite this article as: V. A. Stepanov, N. D. Zhigadlo, E. Piatti and D. Daghero. \href{https://doi.org/10.1063/5.0354116}{\textit{APL Materials} \textbf{14}, 101103 (2026)}.   
\end{abstract}

\maketitle

\section{Introduction}

The lack of an inversion center in the crystal structure of a superconducting material leads to profound consequences on the properties of its superconducting state\,\cite{Takayama2012, Bauer2004, Smidman2017}.
The discovery of heavy-fermion superconductor CePt$_3$Si\,\cite{Bauer2004} sparked a surge of interest in non-centrosymmetric superconductors, when it became apparent that breaking the inversion symmetry in a superconductor could endow it with unconventional properties that are not permitted when inversion symmetry is preserved\,\cite{Smidman2017} and could potentially enable the formation of superconducting pairings with a mixture of spin-singlet and triplet components and other exotic pairing states\,\cite{Smidman2017}.
In 2012, Takayama et al.\,\cite{Takayama2012} discovered a new family of {\color{blue} centrosymmetric} superconducting ternary platinum pnictides, with chemical formula $A$Pt$_3$P with $A=$ Sr, Ca, or La, characterized by a distorted anti-perovskite crystal structure similar to that of the non-centrosymmetric heavy-fermion superconductor CePt$_3$Si\,\cite{Bauer2004}.
This suggested that a comparison of the characteristics of the superconducting state of {\color{blue} non-centrosymmetric} CePt$_3$Si and {\color{blue}centrosymmetric} $A$Pt$_3$P could allow understanding the role of inversion symmetry in the pair formation mechanism\,\cite{Smidman2017}.
Actually, not all $A$Pt$_3$P compounds display the same properties.
For example, LaPt$_3$P has the lowest superconducting critical temperature of the family ($T_c=1.1$\,K) and a different electronic structure with respect to the other two members; it has also been argued to host a chiral $d$-wave singlet superconducting ground state\,\cite{Biswas2021}; the other two compounds, i.e. CaPt$_3$P and SrPt$_3$P, seem to display less exotic properties, but conflicting results appeared in literature still ask for further investigation. 

In particular, the superconductivity of SrPt$_3$P, which displays the highest $T_c\simeq8.4$\,K, was investigated in a number of experimental works on polycrystals.
Measurements of the Hall effect, heat capacity, magnetic field penetration depth $\lambda(T)$, temperature dependence of the upper critical field $H_{c2}$\,\cite{Takayama2012, Khasanov2014} gave grounds to assert that the electronic structure of this compound is multiband, the symmetry of the gap function is $s$-wave, and that the gap values on the two bands are equal ($\Delta_1 = \Delta_2$).
However, Refs.\,\onlinecite{Takayama2012} and \onlinecite{Khasanov2014} gave different values of the energy gap (1.85\,meV and 1.55\,meV at low temperature) and of the gap ratio ($2\Delta/k_\mathrm{B}T_c \approx 5$ and $2\Delta/k_\mathrm{B}T_c = 4.35$, respectively).
NMR measurements of the Knight shift and of the spin-lattice relaxation rate confirmed the $s$-wave symmetry of the pairing and the strong electron-phonon coupling, together with the absence of significant electron correlations\,\cite{Shiroka2015}.

First-principle calculations performed by different groups substantially agree on the electronic band structure, on the existence of multiple bands crossing the Fermi level that give rise to a complicated 3D Fermi surface\,\cite{Nekrasov2012, Chen2012, Subedi2013}, on the negligible role of spin-orbit coupling\,\cite{Chen2012, Subedi2013} and on the existence of low-frequency phonon modes\,\cite{Chen2012, Subedi2013} that are fundamental in determining the electron-phonon coupling\,\cite{Kang2013}.
However, they disagree on various important points, e.g. the existence of a CDW instability (predicted in Ref.\,\onlinecite{Chen2012} but ruled out in Ref.\,\onlinecite{Subedi2013}) that could enhance the coupling strength resulting in $2\Delta/k_\mathrm{B}T_c = 5$, the multiband character of superconductivity (proposed in Ref.\,\onlinecite{Takayama2012} and \,\onlinecite{Chen2012} but denied in Ref.\,\onlinecite{Subedi2013}) or the pairing strength.
In Ref.\,\onlinecite{Subedi2013} a conventional $s$-wave pairing with no gap anisotropy was proposed, as well as a gap ratio $2\Delta/k_\mathrm{B}T_c = 4.06$ perfectly compatible with the fit of the specific-heat transition measured in Ref.\,\onlinecite{Takayama2012} within a single-gap Migdal-Eliashberg model. 
Experimental and theoretical studies of Pd-doped SrPt$_3$P polycrystals\,\cite{Hu2016} highlighted that the mutual influence of electron correlations, electron-phonon coupling, and spin-orbit coupling (SOC) can lead to an increase in the energy gap and non-standard pairing\,\cite{Hu2016}.

More recently, high-quality and sufficiently large SrPt$_3$P single crystals ($T_c = 8.6$\,K) were grown in a high-temperature cubic anvil at high pressure\,\cite{Zhigadlo2016}.
This gave a new impulse to experimental studies of this compound.
In Ref.\,\onlinecite{Cho2022} the anisotropy of $H_{c2}(T)$, of $\lambda(T)$, and of the resistivity $\rho(T)$ were studied for the first time in single crystals{\color{blue}, yielding a pronounced dependence of $\lambda(T)$ on the orientation of the applied magnetic field, at variance with the isotropic behavior displayed by both $\rho(T)$ and $H_{c2}(T)$.
T}he authors interpreted {\color{blue}these results} in terms of a non-standard pairing with point nodes in the order parameter of this compound.

Thus, the question about the pairing strength, of the magnitude and even of the symmetry of the superconducting energy gap in SrPt$_3$P remains open.
In this work, we provide a direct measurement of the energy gap of high-quality SrPt$_3$P single crystals with $T_c = 8.6$\,K by means of Andreev reflection spectroscopy.
Our measurements provide clear evidence for a single isotropic $s$-wave gap in this compound, with a temperature dependence in good agreement with BCS theory, a zero-temperature mean value $\Delta(0)=1.88\pm0.08$\,meV, and a gap ratio $2\Delta/k_\mathrm{B}T_c = 5.2\pm0.2$.

\section{Experimental methods}

Single crystals of SrPt$_3$P were grown under high pressure using a cubic-anvil apparatus and mechanically separated from the solidified flux as described in detail in Ref.\,\onlinecite{Zhigadlo2016}.
The assembled high-pressure cell was compressed to 2\,GPa at room temperature and subsequently heated to 1500\,$^\circ$C over 4\,h.
The maximum temperature was maintained for 20\,h to ensure complete melting and homogenization of the mixture.
The melt was then cooled to 1150\,$^\circ$C over 40\,h to promote crystal growth, followed by cooling to room temperature over 3\,h before releasing the pressure.
The resulting crystals exhibited a pillar-like morphology, as shown in the inset to Figure 1(a).
Single-crystal X-ray diffraction confirmed that SrPt$_3$P crystallizes in the tetragonal P4/nmm space group with lattice parameters $a = b = 5.8\,$\AA\, and $c = 5.4\,$\AA\,\cite{Zhigadlo2016}.
Further details of the crystal growth procedure and characterization are available elsewhere\,\cite{Zhigadlo2016, Cho2022}.

All measurements were performed on single-crystal SrPt$_3$P with typical dimensions of $\approx 1.2\times0.7\times0.2$\,mm$^3$.
The Andreev reflection spectra –- i.e. the dependences of the differential conductance of microcontacts between a superconducting sample S and a non-superconducting metal N on the voltage across the contact, i.e. $\sigma_\mathrm{SN}(V,T) = dI/dV(V)$ -– were measured on “soft” point contacts (PCs) made between the SrPt$_3$P crystal and a small drop of Ag glue, although several control measurements were actually made on mechanically adjustable contacts between the SrPt$_3$P crystals and an Au tip.
The technique of working with such contacts and their characteristics are described in detail in Refs.\,\onlinecite{Gonnelli2002, Daghero2010, Gorina2017}.
PCs were created on the flat side or edge of single crystals that were not subjected to any processing.
As shown in Refs.\,\onlinecite{Zhigadlo2016} and \onlinecite{Cho2022}, the facets of the crystals do not have a simple correspondence with the crystallographic planes.
This, together with the uneven surfaces, does not allow us to draw any strong conclusion about the direction of the current injected into the crystal with respect to the crystal axes.
This creates no problem if the gap is isotropic, but makes a clear observation of the possible anisotropies more challenging.
The high stability of the “soft” PCs made it possible to accurately measure their differential conductance $\sigma_\mathrm{SN}(V,T)$ at different temperatures and to determine the critical contact temperature $T_c^\mathrm{A}$, at which the characteristic structures associated with the Andreev reflection disappear.
More precisely, we will identify $T_c^\mathrm{A}$ as the midpoint between the temperature of the last superconducting curve and the temperature of the first non-superconducting one.
For all contacts, $T_c^\mathrm{A}$ is perfectly compatible with the critical temperature of the crystals measured by magnetic susceptibility\,\cite{Zhigadlo2016}.

The current-voltage ($I-V$) characteristics of the contacts were recorded using a Keithley 6221 current source and a Keithley 2182A nanovoltmeter. The conductances were obtained by numerical differentiation of the $I-V$ characteristics.
The temperature $T$ was measured with a RuO thermometer mounted on the copper head of the cryogenic insert, close to the sample.

The conductances measured at different temperatures $\sigma_\mathrm{SN}(V,T)$ were normalized to the conductance in the normal state $\sigma_\mathrm{N}(V)$ measured at a temperature just (some tenths of a Kelvin) higher than $T_c^\mathrm{A}$.
The normalized conductances $\sigma_\mathrm{n}(V,T) = \sigma_\mathrm{SN}(V,T)/\sigma_\mathrm{N}(V)$ were fitted by using both the original BTK (Blonder, Tinkham and Klapwijk) model\,\cite{Blonder1982} and its extension to the 2D case\,\cite{Tanaka1995, Kashiwaya2000}, always including the broadening parameter $\Gamma$\,\cite{Plecenik1994}.
This made it possible to determine the temperature dependences of the energy gap of the crystal, $\Delta(T)$, as well as the transparency of the potential barrier at the SN interface expressed by a dimensionless parameter $Z$, and the broadening parameter $\Gamma$ that accounts for any non-thermal source of smearing of the curves.
In one case, we also employed the model for Andreev reflection in the presence of critical current effects described in Ref.\,\onlinecite{Daghero2023}.

\section{Results and discussion}

\begin{figure}
\includegraphics[width=\columnwidth]{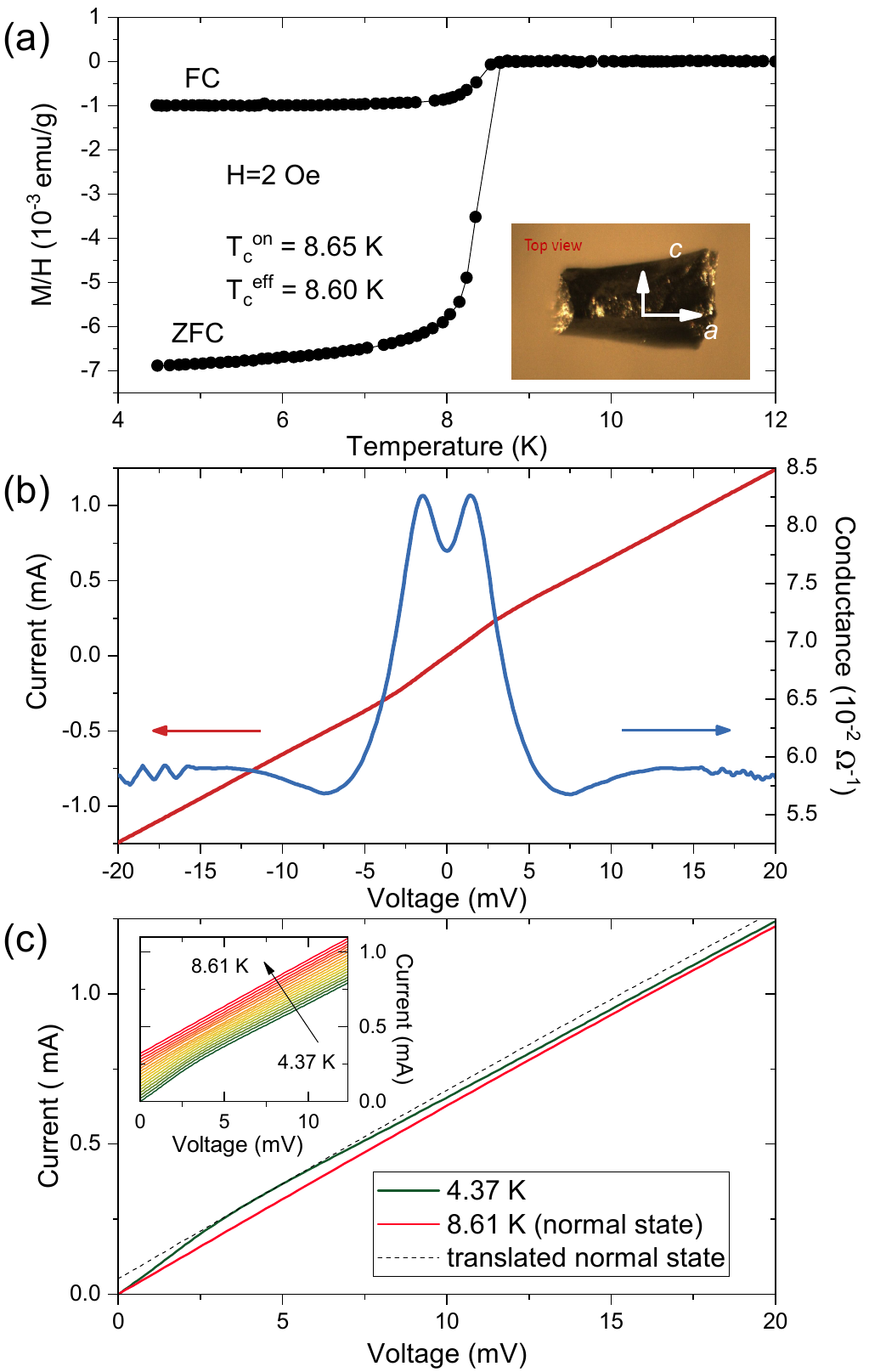}
\caption{\label{fig:1}
(a) Magnetization as a function of temperature for one of the crystals used in this work, measured in field-cooling (FC) and zero-field cooling (ZFC) conditions in a magnetic field of 2\,Oe. The inset shows a picture of a typical crystal with the approximate direction of the crystal axes.
{\color{blue}(b) Current-voltage characteristic ($I-V$, left scale) of a point contact with resistance $R_\mathrm{N}=16.4\,\Omega$ measured at 4.37\,K, and corresponding differential conductance curve ($dI/dV$, right scale). (c) The positive branch of the $I-V$ curve at the lowest temperature, i.e. 4.37 K (green) and of the $I-V$ curve in the normal state, i.e. at 8.61 K (red). The dashed line represents the latter, vertically translated to highlight the progressive disappearance of the excess current in the superconducting curve. The inset shows the positive branch of the $I-V$ curves at various temperatures, from 4.37 K to 8.61 K, vertically offset for clarity. }}
\end{figure}

Figure 1(a) shows the magnetization curve of one of the crystals used in this paper, that clearly displays a superconducting transition at $T_c^\mathrm{eff} = 8.60$\,K.
{\color{blue}In general, the crystals used in the present work were consistent with those described in Refs.\,\onlinecite{Zhigadlo2016} and \onlinecite{Cho2022}.}
{\color{blue}Figure 1(b) displays, as an example, the $I-V$ curve (left scale) of a point contact (later on called PC2) with resistance $R_\mathrm{N}=16.4\,\Omega$, recorded at $T=4.37$\,K, and the relevant conductance $dI/dV$ (right scale). 
The $I-V$ curve clearly shows an excess current at low voltage, i.e. $|V|\lesssim 3$ meV, which is associated to Andreev reflection and results in an enhanced conductance with respect to the high-bias region. 

Figure 1(c) compares the $I-V$ curve in the superconducting state at the lowest temperature (green) to that in the normal state, just above the transition (red). The slope of the two curves is the same at high voltage ($V>15$ mV) and this means that the conductance is the same in this bias range. However, the excess current is progressively lost; if it was retained, the high-bias part of the superconducting curve would follow the dashed line, which has the same slope as the normal-state one, but is vertically offset by the excess current. The progressive suppression of the excess current results in characteristic dips in the differential conductance, clearly seen in Figure 1(b). The origin of these dips will be widely discussed in the following. The inset to Figure 1(c) shows the effect of temperature on the $I-V$ curves, i.e. a progressive weakening of the change in slope at the gap edge, and its complete disappearance at the critical temperature.}

Figure 2 shows differential conductance curves of four “soft” SrPt$_3$P/Ag PCs with resistances $R_\mathrm{N} = 43.5, 16.4, 13.8$ and $2.0\,\Omega$ at different temperatures.
{\color{blue}The first three were obtained on the same crystal mentioned above \cite{Zhigadlo2016,Cho2022}, with $T_c^\mathrm{eff} =8.6$\,K; } the last was measured in another sample, synthesized at lower temperature and ceramic-like, with magnetic critical temperature $T_c^\mathrm{eff} =8.4$\,K.
All conductances are symmetric with respect to the point $V = 0$, feature an enhancement of conductance at low bias typical of Andreev reflection, and a pair of symmetric conductance maxima at $\pm±V^*\simeq\pm(1-1.5)$\,mV usually associated with a full superconducting gap.
With increasing temperature, the peaks move toward lower voltage and at $T \approx 6$\,K they merge in a single broad maximum at zero bias.
The signatures of Andreev reflection are progressively suppressed and the high-bias tails of the curves, that are all superimposed, display a downward curvature which is characteristic of PCs with metallic conductivity\,\cite{Srikanth1992}.
The superposition of the high-bias tails, and the absence of any vertical shift of such tails on increasing temperature (especially in the region of the resistive transition), indicates a negligible spreading resistance\,\cite{Chen2010, Baltz2009, Bugoslavsky2005} associated to the normal state of the crystals\,\cite{Daghero2018, Pecchio2013}.
In other words, the resistance of the junction is much higher than the resistance of the crystal even when the latter is in the normal state.
This ensures that the potential difference between the banks of the point contact correctly represents the excess energy with which electrons impinge on the normal/superconductor interface.
At the critical temperature $T_c^\mathrm{A}$ (indicated in each panel) the Andreev signal disappears.
The curve recorded immediately above $T_c^\mathrm{A}$ is identified with the normal-state conductance of the specific contact, $\sigma_\mathrm{N}(V)$, whose convex shape (with a slight decrease in conductance with increasing $V$) is typical of microshorts in the “diffusion” regime\,\cite{Kulic1977, Naidyuk2005}.
{\color{blue} In particular, the normal-state conductance is linear as a function of $V$ for $|V|\gtrsim 6 $ meV.
More precisely, the second derivative $-d^2 I/dV^2$ shows an abrupt change in slope at $|V| \simeq 6$ mV, and becomes constant above that value, independent of the contact resistance.
This reflects a genuine spectroscopic mechanism, i.e. the excitation of the low-energy phonon modes typical of SrPt$_3$P, which are strongly coupled to electrons and largely contribute to the electron-phonon coupling constant \cite{Subedi2013,Kang2013}.
The linear dependence of $\sigma_\mathrm{N}$ on $V$ for $|V|\gtrsim 6 $ meV (i.e. the saturation in the second derivative $-d^2I/dV^2$) is the hallmark of the crossover to a stable, non-equilibrium "phonon bottleneck" regime \cite{Naidyuk2005,Kulic1977} due to the local accumulation and reabsorption of non-thermal phonons, in which the contact conductance is intrinsically linear with voltage.
This mechanism does not necessarily give rise to significant heating effects, and 
experimentally these are highly unlikely
at least for PC1, PC2 and PC3, because the relative change in $\sigma_\mathrm{N}(V)$ between zero bias and, for instance, $|V|=15$ mV, is tiny: 2\% for PC1, 6-7\% for PC2 and PC3.
If such variations were due to heating effects, they would translate into an increase of temperature by 1 K for PC1, and 4 K in PC2 and PC3 -- one order of magnitude smaller than what expected for a contact in the Maxwell (thermal) regime, i.e. about 40 K.
In the case of PC4 the change in $\sigma_\mathrm{N}(V)$ is about 20\% and the equivalent temperature increase would be 10 K -- anyway excluding a pure thermal regime. }

\begin{figure}[t]
\includegraphics[width=\columnwidth]{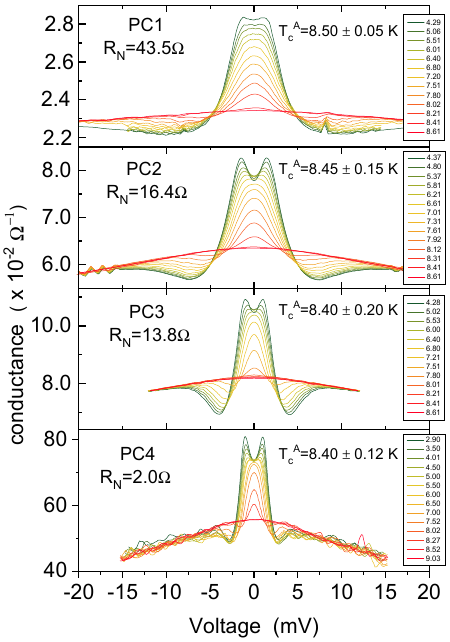}
\caption{\label{fig:2} Differential conductances of four “soft” SrPt$_3$P/Ag point contacts (PC1 to PC4) with normal-state resistances $R_\mathrm{N} = 43.5, 16.4, 13.8$ and $2.0\,\Omega$ at different temperatures. On decreasing the contact resistance, the conductance minima shift towards lower energies, as expected for dips due to the attainment of the critical current in the contact region, partially eroding the signatures of the superconducting gap.}
\end{figure}

The conductance minima (dips) at finite voltage $\pm V_\mathrm{dip}$ that are clearly visible in all the panels are instead an anomalous feature that is worth a discussion.
First of all, these minima occur at bias voltages that vary from contact to contact, and are thus not related to intrinsic properties of the material under study.
Rather, the position of the minima scales with the contact resistance, indicated in each panel: the smaller is the normal-state resistance, the smaller is the voltage corresponding to the minimum.
This clearly indicates that the minima are likely to occur at a specific current rather than to a specific voltage (or energy).
It is also worth noting that the minima are fairly sharp in low-resistance contacts like PC4 and PC3 but are instead rather wide (extending up to 20\,mV) in the most resistive ones, i.e. PC1 and PC2.
On increasing the temperature, the depth of the dips and their width decreases, while they move to lower voltages.
The presence of these dips, as we already discussed in the case of Mo$_3$Al$_2$C\,\cite{Zhigadlo2018}, may in principle interfere with the determination of the gap amplitude while fitting the experimental spectra, and thus requires particular caution in the analysis of the data. 

As for the origin of these dips, let us recall first that when measuring the energy gap using the Andreev reflection method, it is necessary to ensure that the contacts are in the spectroscopic regime\,\cite{Daghero2010, Chen2010, Naidyuk2005, Deutscher2005}.
For this, the characteristic contact size $d$ must be much smaller than the mean free path of electrons in the two banks, $d \ll l_i,l_e$ (ballistic mode) or, at least, smaller than the energy relaxation length of electrons $\sqrt{l_\mathrm{i}l_\mathrm{e}}$ (diffusion mode), where $l_\mathrm{i}$ and $l_\mathrm{e}$ are the inelastic and elastic mean free paths of electrons.
In these conditions, the size of the contact can be estimated by starting from the contact resistance through Sharvin’s formula\,\cite{Sharvin1965} or Wexler’s formula\,\cite{Wexler1966} provided that the mean free path is known.
In our N/S contacts, the most stringent limit on the contact size $d$ is posed by the values of $l_\mathrm{i}$ and $l_\mathrm{e}$ in the SrPt$_3$P crystal, since its resistivity ($\rho_0 \approx 7\,\upmu\Omega$\,cm\,\cite{Cho2022}) is certainly greater than that of high-purity Ag ($\rho_0 \approx 1.3\,\upmu\Omega$\,cm).
Unfortunately, the mean free path in SrPt$_3$P 
{\color{blue}single crystals grown by the high-pressure high-temperature method\,\cite{Zhigadlo2016, Cho2022}} 
is unknown{\color{blue}, and accurately extracting it from transport measurements is extremely challenging due to the complicated Fermi surface\,\cite{Nekrasov2012, Chen2012, Subedi2013} comprising multiple inequivalent pockets that result in a strongly non-linear Hall coefficient\,\cite{Takayama2012}.
The mean free path of SrPt$_3$P crystals grown via the Bridgman method has been reported to range between 38 and 59\,nm\,\cite{Horie2022}, but these results cannot be directly translated to our crystals due to the much higher resistivity of the Bridgman crystals and the incomplete mapping of all carrier branches\,\cite{Horie2022}.}
Therefore, we cannot {\color{blue}reliably} estimate the transport mechanism through the PC based on its resistance.
Fortunately, hints on the conduction regime through the PC can be obtained by looking at the conductances $\sigma_\mathrm{SN}(V,T)$, recorded at different temperatures up to $T_c^\mathrm{A}$.
In this sense, the presence of the dips certainly indicates that the contacts are not “ideal”, but does not necessarily rule out the possibility of extracting the gap amplitude from the fit of the experimental curves.

There are indeed various possible origins of these features, that are often observed in point-contact spectroscopy.
The proximity effect\,\cite{Strijkers2001} cannot be invoked here because it would make dips appear at voltages smaller than the bulk gap (while here $V_\mathrm{dip}$ can be as high as 10\,mV).
The existence of intergrain Josephson junctions (weak links) in series with the contact\,\cite{Shan2003, Lee1971} is highly unlikely as well, due to the crystalline (and very hard) nature of the crystals.
Another possible source of dips is the existence of a degraded, non superconducting layer $n$ of thickness $d_n$ on the surface of the sample\,\cite{Zhitlukhina2016}, as it may happen because of oxidation.
In the resulting N-I-$n$-S junction, electrons can be trapped in the $n$ layer because of Andreev reflection at the $n$-S interface and normal reflection at the I-$n$ interface.
The energy of the resulting bound states, which is equal to the bulk superconducting gap $\Delta$ in the absence of the surface layer, is reduced by an amount that strongly depends on the thickness $d_n$.
Then, the coherence peaks of the tunnel conductance are shifted to lower voltage ($V^*<\Delta/e$) while a dip appears slightly above $V^*$\,\cite{Zhitlukhina2016}.
Clearly, such a shift prevents a proper determination of the bulk gap amplitude.
This mechanism, however, seems not to be able to explain our results because the dips are expected to be narrow and appear at a voltage $V_\mathrm{dip}\leq\Delta/e$ while in our curves the dips occur at a very high voltage and are, moreover, very wide. 

The last and more reasonable explanation for the dips observed in our spectra is the suppression of superconductivity in the contact region, caused by the transport current flowing through the contact{\color{blue}, as originally proposed in Ref.\,\onlinecite{Sheet2004}}.
When the current density exceeds a critical value, a local transition of the superconductor to the resistive state occurs\,\cite{Haussler1996, Naidyuk1991}.
If the contact is not perfectly in the ballistic regime, its resistance is described by Wexler’s formula\,\cite{Wexler1966}
\begin{equation}
    R = \frac{8h}{e^2d^2k_\mathrm{F,min}\tau}+\gamma\left(\frac{l}{d}\right)\frac{\rho_1+\rho_2}{2d}
    \label{eq:Wexler}
\end{equation}
as the sum of the Sharvin term and of a Maxwell term that is negligible as long as the resistivity of the material $\rho_1$ is zero, since the resistivity $\rho_2$ of the normal tip is usually very small.
However, the sudden onset of a non-zero $\rho_1$ gives rise to an increase in resistance (and a decrease in conductance) which is detected as a dip in the PC spectra\,\cite{Daghero2010, Sheet2004, Daghero2023}.
The current at which the dip occurs is fixed by the critical current density (i.e. by the material) and by the size of the contact, but the corresponding bias voltage varies depending on the contact resistance, as observed experimentally.  

{\color{blue} The onset of dips can (though not necessarily) lead to dissipation in the contact, and thus to local heating effects\,\cite{Naidyuk2005, Daghero2006}. Here, the coincidence of the high-bias tails of the conductance curves at any $T<T_c^A$ with the normal-state ones $\sigma_\mathrm{N}(V,T)$ at $\left|V\right| > (10-15)$\,mV (Figure 2) indicates that, once the critical current is exceeded, the system simply reverts to the same non-equilibrium phonon bottleneck regime that governs the normal state. Even if heating effects were present, they would be marginal and, especially, they would set in jointly with the appearance of the dips, without affecting the spectra at lower bias. This is the reason why, as already pointed out, $T_c^\mathrm{A}\approx T_c$.} 

The shape of the dips reflects the way in which the material transitions to the normal state, i.e. the shape of the resistivity vs. current curve of the material\,\cite{Sheet2004}.
In Ref.\,\onlinecite{Daghero2023} we proposed a model for the critical current effects in PCs that allows reproducing the whole spectrum, including the dips, by using a general and rather versatile equation to describe this curve.
This method has the advantage of partially correcting for the effect of the dips, even when they occur rather close to the gap features in the spectra, as shown in the case of MgAlB$_2$\,\cite{Daghero2023}.
If the current-induced appearance of a Maxwell resistance starts at a bias voltage $V_\mathrm{crit} > V_\Delta = \Delta/e$ ($\Delta$ being the energy gap of the superconductor, $e$ the electron charge) all the region of the spectra at $\left|V\right|<V_\mathrm{crit}$ can be still considered reliable, and it is possible to extract the gap amplitude by means of a conventional fit with the models for Andreev reflection\,\cite{Blonder1982, Tanaka1995, Kashiwaya2000} as demonstrated in literature\,\cite{Bashlakov2005, Sirohi2016}, at least at low temperature.
It is important to note that the decrease in the critical current on increasing temperature makes the dips move to lower bias so that they could end up preventing the correct extraction of the gap amplitude, as shown in Refs.\,\onlinecite{Zhigadlo2018} {\color{blue}and \onlinecite{Sheet2004}}.

If instead $V_\mathrm{crit}< V_\Delta$, the structure of the Andreev signal can change completely and it becomes impossible to extract the gap amplitude from the spectra.
For example, in the case of a $s$-wave gap, the conductance peaks may no longer be determined by the energy gap, but by the voltage $V_\mathrm{crit}$; in extreme cases, only a zero-bias conductance peak surrounded by dips survives, which makes the spectra look more similar to what is expected with a nodal symmetry (e.g. $d$-wave)\,\cite{Sheet2004}. 

Going back to the spectra shown in Figure 2, the presence of a characteristic double-maximum structure is certainly suggestive of a full gap, with no nodes.
However, in principle one is not sure whether the position of these maxima is set by the gap itself (and is thus intrinsic) or by $V_\mathrm{crit}$ (and is thus extrinsic).
Reasonably, one expects PC1 and PC2 (that display the peaks at the maximum voltage, and have larger resistance) to be more probably spectroscopic, while in the spectra of contacts PC3 and PC4 the position of the maxima is clearly reduced by the dips.
In any case, to check if the maxima are related to the energy gap of the superconductor, it is necessary to fit the conductances measured at different temperatures with the functions calculated according to the BTK theory\,\cite{Blonder1982} or its generalizations\,\cite{Tanaka1995, Kashiwaya2000}, find $\Delta(T)$ and make sure that this dependence, at least partially, coincides with the BCS dependence\,\cite{Naidyuk2005, Zhigadlo2018}.

To fit the measured conductances with theoretical functions, the former must be normalized to the contact conductance in the normal state $\sigma_\mathrm{N}(V)$, recorded when both the banks of the junction are in the normal state.
A magnetic field or temperature can be used for this.
Using a magnetic field would allow us to measure $\sigma_\mathrm{SN}(V,T)$ and $\sigma_\mathrm{N}(V)$ at the same temperature and eliminate the possible change in $\sigma_\mathrm{N}(V)$ with temperature.
However, the change in $\sigma_\mathrm{N}(V)$ with temperature in the range of $(4-10)$\,K is insignificant and the $\sigma_\mathrm{N}(V)$ curves obtained by suppressing superconductivity either by a magnetic field or temperature are expected to coincide. 
As already pointed out before, we took the conductances recorded at $T$ slightly exceeding $T_c^\mathrm{A}$ as $\sigma_\mathrm{N}(V)$.
After normalization of $\sigma_\mathrm{SN}(V,T)$, all the above-mentioned features on the dependences $\sigma_\mathrm{n}(V,T)= \sigma_\mathrm{SN}(V,T) / \sigma_\mathrm{N}(V)$ were preserved.

\begin{figure}
\vspace{0.85mm}
\includegraphics[width=\columnwidth]{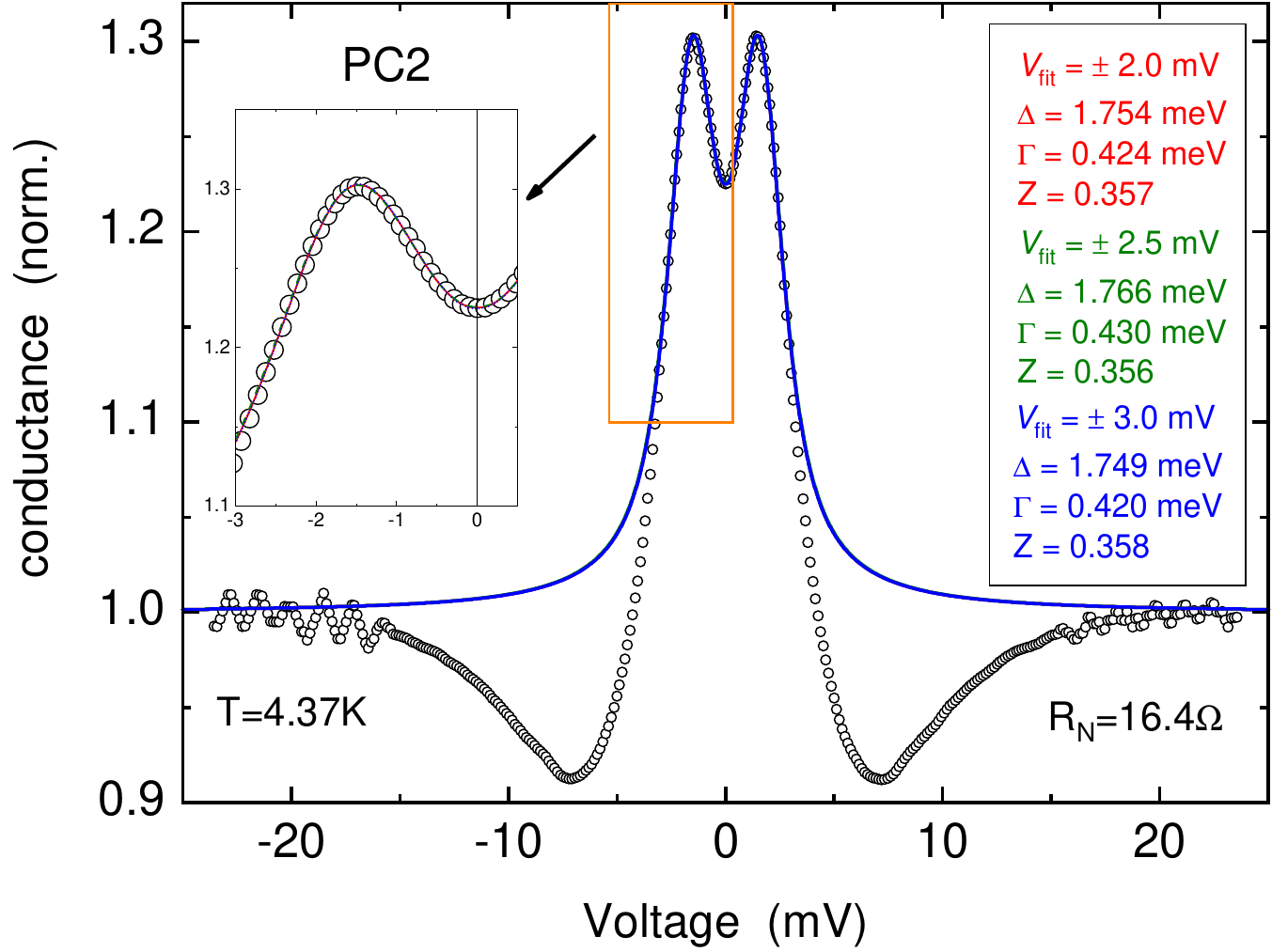}
\caption{\label{fig:3} Differential conductance of PC2, recorded at $T = 4.37$\,K (symbols), normalized to the curve at 8.6\,K, together with the relevant best-fitting curves (lines) obtained by an automatic least-square fitting procedure restricted to different energy ranges: $\pm2.0$\,meV (red), $\pm2.5$\,meV (green) and $\pm3.0$\,meV (blue). The inset shows a magnification of the region of the maximum and highlights the extremely good agreement between experiment and calculations up to 2.5\,meV, as well as the practical indistinguishability of the three fitting curves in this energy scale.}
\end{figure}

Figure 3 shows the normalized conductance $\sigma_\mathrm{n}(V)= \sigma_\mathrm{SN}(V) / \sigma_\mathrm{N}(V)$ of PC2 with $R_\mathrm{N} = 16.4\,\Omega$ at a temperature of 4.37\,K.
None of the usual models for Andreev reflection is able  the reproduce the $\sigma_\mathrm{n}(V)$ curve in the entire voltage range, but we can try to identify the bias voltage $V_\mathrm{crit}$ at which the superconducting state starts to be suppressed in the contact region.
To do so, we fitted the experimental $\sigma_\mathrm{n}(V)$ shown in Figure 3 with the 2D-version of the BTK model for Andreev reflection in a $s$-wave superconductor\,\cite{Tanaka1995, Kashiwaya2000}, in a restricted voltage range $\left|V_\mathrm{fit}\right| = 1.0$\,mV, 1.1\,mV, 1.2\,mV... etc.
The fit was performed by an automatic least-square procedure.
After finding the values of the parameters $\Delta$, $\Gamma$ and $Z$ that allow best fitting $\sigma_\mathrm{n}(V)$ in the range $\left|V_\mathrm{fit}\right|$, the theoretical curve was calculated in the entire voltage range.
The accuracy of the fit in the $whole$ range was estimated by looking at how it could reproduce the position of the maxima, the zero-bias minimum as well as the overall shape of the curve within the maxima.
The best results were obtained when fitting $\sigma_\mathrm{n}(V)$ in the voltage range $\left|V_\mathrm{fit}\right| = (2-2.5)$\,mV.
Figure 3 shows the three theoretical curves obtained with $\left|V_\mathrm{fit}\right|=2.0, 2.5$ and $3.0$\,mV which coincide almost perfectly up to 2.5\,mV, but slightly differ at intermediate voltages (with a maximum difference around 5\,mV, not detectable in the figure), and finally become again indistinguishable at $V\simeq20$\,mV.

\begin{figure}
\includegraphics[width=0.96\columnwidth]{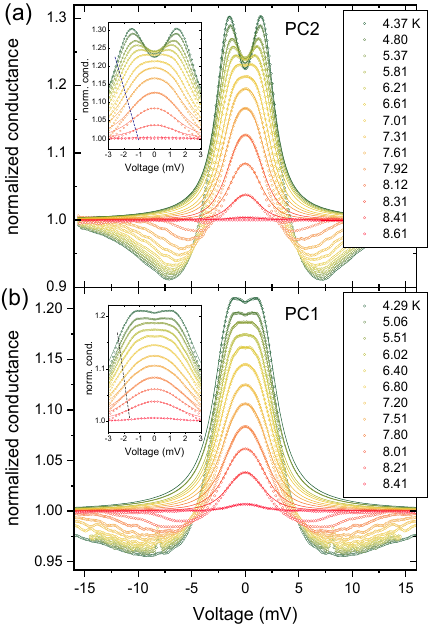}
\caption{\label{fig:4}
(a) Normalized conductance $\sigma_\mathrm{n}(V,T)$ of PC2 with $R_\mathrm{N} = 16.4\,\Omega$ in the temperature range $T = (4.3-8.6)$\,K (symbols).
The lines show the normalized conductances $\sigma(V,T)$ calculated using the two-dimensional version of the BTK theory with $s$-symmetry of the energy gap\,\cite{Tanaka1995, Kashiwaya2000}, fitted to the measured values using the criterion of the minimum sum of squared deviations in a temperature-dependent range, from $\left|V_\mathrm{fit}\right| = 2.0$\,mV at 4.37\,K, to $\left|V_\mathrm{fit}\right| = 1$\,mV close to $T_c^\mathrm{A}$.
The inset shows the conductances $\sigma_\mathrm{n}(V,T)$ in the range $|V|\leq3$\,mV.
The straight dashed line approximately indicates the limit voltage up to which the fit was restricted.
(b) The same for PC1 with $R_\mathrm{N} = 43.5\,\Omega$.
}
\end{figure}

This method just tells us that, within $\pm2.5$\,mV, the experimental spectrum can be reproduced by a theoretical model that does not account for the dips, but it does not guarantee that the obtained value of $\Delta$ is not distorted and really corresponds to the energy gap of the superconductor.
To verify this point, it is necessary to measure the dependence $\Delta(T)$ and check its compliance with the known BCS curve $\Delta_\mathrm{BCS}(T)$\,\cite{Muhlschlegel1959, Poole2014}.
To do so, we fitted the normalized conductance measured at different temperatures shown in Figure 4(a).
Owing to the decrease in $V_\mathrm{crit}$ induced by increasing temperature, we smoothly decreased the range of voltages in which the fitting is carried out.
The inset in Figure 4(a) shows the upper part of the conductances $\sigma_\mathrm{n}(V,T)$ in the range $\left|V\right| \leq3$\,mV.
The straight dashed line approximately indicates the maximum voltages at which the fit was performed.
The best-fitting curves are shown in Figure 4(a) by solid lines; the results of the same procedure applied to the conductance curves of PC1 are shown in Figure 4(b).
As in the previous case, the best fit of the low-voltage part of the lowest-temperature curve was obtained by restricting the fitting range to $\left|V_\mathrm{fit}\right|=2$\,mV.
On increasing temperature, such range was progressively reduced down to $\left|V_\mathrm{fit}\right|=1.5$\,mV.
The inset shows a magnification of the experimental and theoretical curves in the range $\left|V\right|\leq 3$\,mV.

\begin{figure}
\includegraphics[width=0.9\columnwidth]{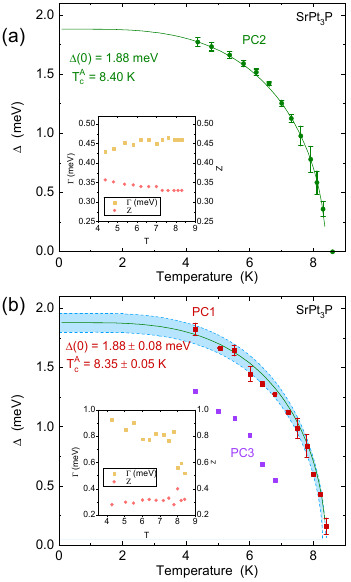}
\caption{\label{fig:5}
(a) Temperature dependence of the parameter $\Delta$ from PC2 (green symbols).
Error bars represent the spread of $\Delta$ values obtained with different models (1D or 2D BTK) and different algorithms for least-square optimization.
The solid line is the $\Delta_\mathrm{BCS}(T)$ line that best approximates the data, with $\Delta_0=1.88$\,meV and $T_c^\mathrm{A}=8.4$\,K.
The inset reports the other fitting parameters $\Gamma$ and $Z$.
(b) $\Delta(T)$ from PC1 (red symbols).
The light blue region is bounded by two BCS-like curves with $\Delta_0=1.96$\,meV, $T_c^\mathrm{A}=8.4$\,K and $\Delta_0=1.80$\,meV, $T_c^\mathrm{A}=8.3$\,K.
The inset show the other fitting parameters (referred to PC1).
Violet symbols in the main panel represent the parameter $\Delta$ extracted from PC3, which is clearly suppressed by dips and could (at most) be poorly compatible with a BCS-like trend with reduced $\Delta_0=1.44$\,meV and reduced $T_c^\mathrm{A}=7.1$\,K.
}
\end{figure}

Figure 5(a) reports the temperature dependences of the parameter $\Delta$ obtained from the fit of the conductance curves of contact PC2 (symbols).
In order to estimate the uncertainty of these values, we used two different models for the fit (the 2D BTK model\,\cite{Tanaka1995, Kashiwaya2000}, whose theoretical curves were shown in Figure 4, and the 1D BTK model\,\cite{Blonder1982}), and different algorithms for the automatic minimization of the sum of least squares.
The data points shown in Figure 5(a) represent the mean value of the interval of gap amplitudes, the latter being represented as a vertical error bar.
The solid green line is a function of the form
\begin{equation}
\Delta(T)=\Delta_0\tanh \left\{ 1.82\left[1.018\left(\frac{T_c^\mathrm{A}}{T}-1\right)\right]^{0.51} \right\}
    \label{eq:gapBCS}
\end{equation}
due to M{\"u}hlschlegel\,\cite{Muhlschlegel1959}, that accurately approximates the BCS-like $\Delta(T)$ curve.
In particular, the best approximation to the experimental $\Delta(T)$ is obtained with $T_c^\mathrm{A}=8.4$\,K and $\Delta_0=1.88$\,meV.
Note that $T_c^\mathrm{A}$ coincides (within our uncertainty) with the Andreev critical temperature indicated in Figure 2, i.e. $8.45\pm0.15$\,K, as well as with the effective critical temperature measured by magnetization.
It is clear that the values of $\Delta$ extracted from the fit of the spectra of contact PC2 follow almost perfectly a BCS-like trend in the whole temperature range.
This clearly indicates that, in contact PC2, the dips occur sufficiently far from the gap edges not to affect the position or the width of the spectral features associated with the gap itself.
As a matter of fact, if it were not so, an anomalous, almost linear decrease in $\Delta(T)$ -- very different from the BCS trend -- would be observed (as shown for example in low-resistance contacts on Mo$_3$Al$_2$C in Ref.\,\onlinecite{Zhigadlo2018}).
Further confirmation comes from the temperature dependence of the other fitting parameters $Z$ and $\Gamma$ shown in the inset to Figure 5(a).
The values of the two parameters (that are completely unconstrained in the least-square method) are almost constant (note that the vertical scale is very narrow) as is expected for a spectroscopic point contact.
Also note that $\Gamma < \Delta$, as it should happen in good spectroscopic contacts, up to 8.2\,K. 

Figure 5(b) reports the temperature dependence of $\Delta$ from PC1 (red symbols) and PC3 (violet symbols).
As far as PC1 is concerned, note that: i) below 6\,K, the data are more scattered than in PC2, but are compatible with the green curve and thus coincide, once the experimental error bars are taken into account, with those found in PC2.
This equality, despite the large difference in the contact resistances, is a proof that \textit{these values are intrinsic}, i.e. they do not depend at all on the occurrence of dips;
ii) above 6\,K, the gaps of PC1 seem to collapse on a BCS curve that corresponds to a smaller $\Delta_0=1.80$\,meV and a smaller $T_c^\mathrm{A}=8.3$\,K.
It is thus possible that, here, the dips start to have a slight effect on the superconducting features.
Another suggestion in this sense comes from the other fitting parameters, $\Gamma$ and $Z$, shown in the inset to Figure 5(b).
Indeed, when the dips start to erode the superconducting features, the overall enhancement of conductance due to Andreev reflection looks narrower and too sharp to be compatible with the values of $\Gamma$ and $Z$ at low temperature.
Therefore, $\Gamma$ has to be decreased to fit the experimental curves.
This unphysical effect, that indeed happens in PC1, tells us that something extrinsic is at play here.
Despite this fact, the gap amplitudes remain in the colored region which is bounded by two BCS-like curves calculated with $\Delta_0=1.96$\,meV , $T_c^\mathrm{A}=8.4$\,K  and $\Delta_0=1.80$\,meV, $T_c^\mathrm{A}=8.3$\,K.
Putting the results of PC2 and PC1 together, we can conclude with confidence that the low-temperature gap in SrPt$_3$P single crystals is $\Delta_0=1.88\pm0.08$\,meV and the gap ratio is $2\Delta/k_\mathrm{B}T_c=5.2\pm0.2$.

Just for the sake of discussion, let us mention the dataset obtained from PC3, shown in Figure 5(b) as violet symbols.
PC3 has a resistance $R_\mathrm{N} = 13.8\,\Omega$ and a critical temperature that is difficult to identify precisely if one looks at the spectra of Figure 2 due to the presence of various curves that are neither easily identified with “still superconducting” nor as “already normal”.
We can just say that $T_c^\mathrm{A}$ falls somewhere between 8.21 and 8.61\,K, i.e. $T_c^\mathrm{A}=8.41\pm0.20$\,K.
Qualitatively, the raw conductance curves of PC3 do not appear very different from those of PC1 and PC2.
However, one immediately realizes that the parameter $\Delta$ extracted from the fit of these curves (violet symbols in Figure 5(b)) is $not$ representative of the actual superconducting gap because, as a function of temperature, it displays a radically different trend, with a faster than expected decrease on increasing temperature.
This is due to the fact that the superconductivity of this contact begins to collapse at a bias voltage $\left|V_\mathrm{crit}\right| < \left|V_\Delta\right|$ already at low temperature, so that the position of the conductance maxima is determined by the dips rather than by the gap. 
The temperature dependence of $\Delta$ from PC3 is actually \textit{not even comparable} with a BCS-like curve since it rather correlates to the trend of the critical current.
Just for the sake of discussion, the BCS-like curve that would better describe it requires $\Delta_0=1.44$\,meV and $T_c^\mathrm{A}=7.1$\,K, both strongly suppressed with respect to the real ones.

As a further test of the role played by the critical current in determining the shape of the spectra, it is instructive to fit them using the model we recently developed\,\cite{Daghero2023} which is based on the extension of the 2D-BTK model\,\cite{Tanaka1995, Kashiwaya2000} to account for the critical current effects\,\cite{Sheet2004}.
Starting from eq.\,\eqref{eq:Wexler}, we treat the contact as the sum of two terms: one is responsible for Andreev reflection and is modeled by the 2D-BTK model, and the other is a Maxwell term that appears only above a certain critical current (we neglect the contribution of the normal counterelectrode) giving rise to a series resistance which is not constant, but depends on the current, according to the function
\begin{equation}
R_\mathrm{M}=\frac{1}{A\left[\left(\frac{I_c}{I}\right)^{2n}-\left(\frac{I_c}{I}\right)^n\right]+B}
    \label{eq:dip}
\end{equation}
that mimics the $R(I)$ behavior of a wide class of materials, by properly choosing the adjustable parameters $A$, $B$, $I_c$ and $n$\,\cite{Daghero2023}.
In fact, $B$ is fixed by the high-voltage conductance of the contact\,\cite{Daghero2023} so that only the ratio $A/B$ is an adjustable parameter.
The fit of the point-contact spectra with this model has the advantage of not requiring a normalization, i.e. the raw spectra are directly fitted.

\begin{figure}[t]
\includegraphics[width=0.98\columnwidth]{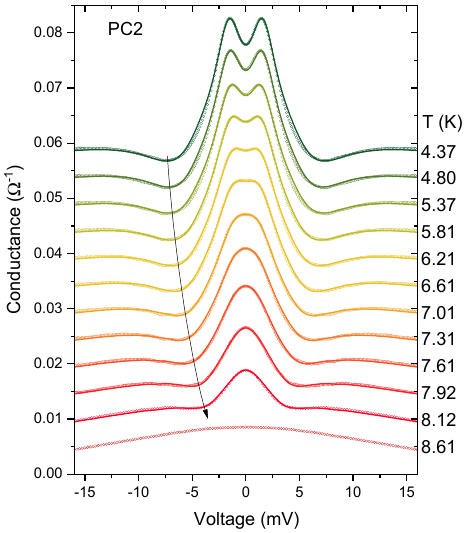}
\caption{\label{fig:6}
The unnormalized (as-measured) differential conductance curves of PC2 (symbols) with their fit obtained by using the 2D BTK model including the effects of the critical current (solid lines).
All the curves apart from the lowest-temperature ones are offset by $-0.005\,\Omega^{-1}$ for clarity.
The curved arrow approximately indicates the displacement of the dips, which corresponds to the decrease of the critical current on increasing the temperature. 
}
\end{figure}

\begin{figure*}[t]
  \centering
  \begin{minipage}[c]{0.65\textwidth}
    \centering
    \includegraphics[width=\linewidth]{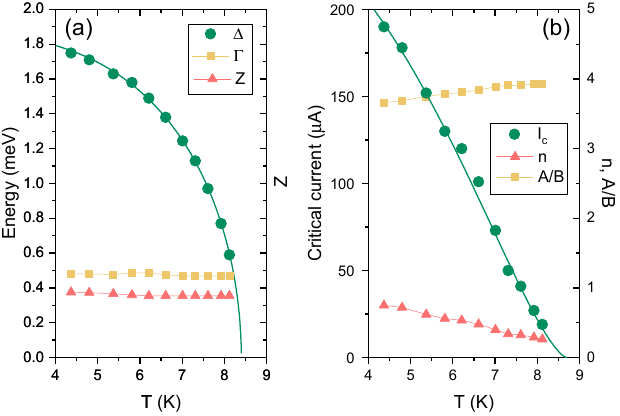}
  \end{minipage}%
  \hfill
  \begin{minipage}[c]{0.3\textwidth}
    \caption{\label{fig:7}
            Temperature dependence of the fitting parameters of the 2D BTK model including the critical current effect.
            (a): the gap amplitude $\Delta$, the broadening parameter $\Gamma$ and the dimensionless barrier parameter $Z$.
            Note that the latter are almost constant.
            The solid line is a BCS-like dependence with $\Delta(0) = 1.87$\,meV and $T_c^\mathrm{A} = 8.4$\,K.  (b): the critical current values used in the fit (green symbols) and the values of the parameters $n$ and $A/B$ that control the shape of the dips.
            The solid line is a temperature dependence of the critical current as taken from Ref.\,\onlinecite{Poole2014} with $I_c(0)=0.265$\,mA and $T_c= 8.7$\,K. 
            }
  \end{minipage}
\end{figure*}

Figure 6 shows the experimental conductance curves of PC2 (the same as in Figure 2) as symbols, together with their best fit obtained with this model (solid lines).
The agreement is impressive in almost all the voltage range and at all temperatures up to $T_c^\mathrm{A}$.
The model follows very well the shape and position of the dips, and allows extracting the temperature dependence of the critical current in addition to that of the energy gap.


Figure 7 (a) reports the temperature dependence of the fitting parameters $\Delta$, $\Gamma$ and $Z$ which is very similar to that obtained with the 2D-BTK fit alone –- if possible, $\Delta$ follows even more accurately the theoretical $\Delta_\mathrm{BCS}$ curve with $\Delta_0=1.87$\,meV and $T_c^\mathrm{A}=8.4$\,K, in perfect agreement with Figure 5(a).
Figure 7(b) shows the values of $I_c$, $n$ and $A/B$.
The critical current follows rather well the trend reported in Ref.\,\cite{Poole2014}, with $I_c (0)=0.265$\,mA and $T_c=8.7$\,K (that must be compared to the onset critical temperature, $T_c^\mathrm{on}=8.65$\,K, measured by susceptibility in this same crystal).
The values of $n$ and $A/B$ look correlated, since they display opposite trends as a function of temperature, as often happens\,\cite{Daghero2023}.
The almost identical trend of $\Delta$, $\Gamma$ and $Z$ as a function of temperature obtained with the sole 2D-BTK model or with the model that accounts for the dips is a further proof that in PC2 the dips systematically occur at energies sufficiently larger than the gap amplitude.
Moreover, the temperature dependence of the critical current is absolutely plausible and demonstrates the reliability of the approach.

\begin{figure}[b]
\includegraphics[width=\columnwidth]{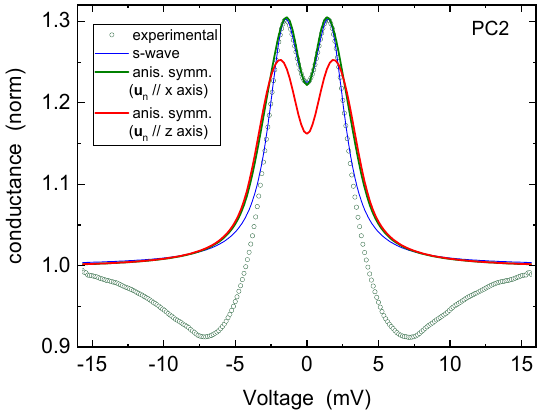}
\caption{\label{fig:8}
A tentative fit of the low-temperature spectrum of PC2 with the anisotropic symmetry proposed in Ref.\,\onlinecite{Hu2016}.
The green line has been obtained with the gap function $\Delta(\theta,\varphi)=\Omega[\sin(\theta)-0.45\cos(2\varphi)]$ using $\Omega=2.5$\,meV, $\Gamma=0.12$\,meV, $Z=0.33$, and assuming that the normal to the S/N interface $\textbf{u}_n$ is along the $x$ axis.
The red curve has been obtained with the same parameters, but assuming that $\textbf{u}_n$ is along the $z$ axis. 
}
\end{figure}

A final discussion must be devoted to the symmetry of the order parameter.
While early measurements of the heat capacity, penetration depth and upper critical field\,\cite{Takayama2012, Khasanov2014} all concurred towards an isotropic, nodeless gap scenario, recently\,\cite{Cho2022} the existence of point nodes has been proposed to specifically account for the dependence of the low-temperature London penetration depth on the magnetic field orientation.
The symmetry proposed is of the form $\Delta(\theta,\varphi)=\Omega[\sin(\theta)-0.45\cos(2\varphi)]$ (where $\theta$ is the colatitude and $\varphi$ is the azimuthal angle) that is anisotropic both in the $xy$ plane and in the $xz$ plane, and features a line of nodes along the $z$ axis.
We tested the compatibility of this scenario with our Andreev spectra by inserting this exact symmetry in our code for the fit of the curves, which yielded a certain variation in the overall spectrum depending on the direction of current injection.
As shown in Figure 8, the low-temperature spectrum of PC2 can be reproduced by this nodal model, at least in the low-voltage range, using the parameters $\Omega=2.5$\,meV, $\Gamma=0.12$\,meV and $Z=0.33$ and assuming that the normal to the S/N interface is along the $x$ axis.
The same parameters give instead a different spectrum, with maxima at higher voltage, if the normal is supposed to be along the $z$ axis.
It must however be pointed out that, for either direction, the nodal model predicts broader structures than in the $s$-wave case (clearly, because of the fact that the gap does not have a single value in all directions).
Most importantly, the spectral broadening introduced by the nodal model clearly compromises the agreement between the BTK model and the experimental Andreev spectra already at voltages of $\pm±2.5$\,mV, i.e. in the crucial voltage range where the shape of the spectrum is not yet affected by critical current effect and is thus most robust.
Our Andreev reflection measurements thus clearly favor the isotropic nodeless scenario\,\cite{Takayama2012, Khasanov2014, Subedi2013} over the nodal one\,\cite{Cho2022}.

\section{Conclusions}

SrPt$_3$P is an interesting system since it displays a distorted anti-perovskite crystal structure similar to non-centrosymmetric heavy-fermion CePt$_3$Si\,\cite{Bauer2004}, but possesses an inversion symmetry center.
Investigation of the superconducting pairing symmetry has been conducted by means of various techniques, i.e. measurements of specific heat, Hall effect, London penetration depth, upper critical field.
Most of the results have been interpreted in terms of a $s$-wave superconducting order parameter, but some disagreement exist on the actual value of the gap amplitude, on the strength of the pairing and also on the multiband/multigap nature of superconductivity.
Here, we have performed point-contact Andreev-reflection spectroscopy on high-quality single crystals of SrPt$_3$P, in order to obtain a direct measurement of the energy gap.
The spectra systematically show pairs of symmetric conductance maxima that are suggestive of a full gap, and no zero-bias peaks (which excludes nodal symmetries with a change of sign of the order parameter).
They also present conductance dips that are due to the current-induced breakdown of superconductivity in the contact region.
When the contact resistance is large enough, these dips fall sufficiently far from the gap energy not to hinder the extraction of the gap amplitude from the spectra by means of a fit with the BTK model or its extended versions.
Our conclusion is that the energy gap is $s$-wave, with an almost perfect BCS-like temperature dependence.
Its low-temperature value is $\Delta_0=1.88\pm0.08$\,meV and the gap ratio $2\Delta/k_\mathrm{B}T_c=5.2\pm0.2$ places this compound in the strong-coupling limit.
The reason of this strength could be, as reported in literature, the existence of low-energy phonons strongly coupled to the electrons.
No evidence of multiple gaps with different amplitude is observed.
The recently proposed anisotropic symmetry (with a line node along the $z$ axis) does not appear compatible with our results, as it would lead to wider spectra than those observed.
Stronger evidence against (or for) the possible validity of the nodal picture calls for the synthesis of crystals with extremely well-oriented facets, as it would allow to probe experimentally the slight differences in the gap amplitudes along the different directions of current injection predicted by the nodal model. 

\begin{acknowledgments}
V.A.S. expresses sincere gratitude to the staff of the Solid State Physics Division of FIAN for their assistance in carrying out this work.
All the authors thank R.S. Gonnelli for perusing the manuscript and for fruitful scientific discussions.
\end{acknowledgments}

\section*{Data Availability Statement}

The data that support the findings of this study are available within the article. Further supporting data are available from the corresponding author upon reasonable request.

\bigskip

\nocite{*}
\bibliography{bibliography}

\end{document}